\documentstyle{article}

\newcommand{\beq}{\begin{equation}}
\newcommand{\eeq}{\end{equation}}

\author{Sergei V. Sushkov\thanks{E-mail: sushkov@kspu.ksu.ras.ru}\\
{\em Department of Geometry, Kazan State Pedagogical University,}\\
{\em Mezhlauk str., 1, Kazan 420021, Tatarstan, RUSSIA}}
\title{\Large Chronology Protection and Quantized Fields:
Complex Automorphic Scalar Field in Misner Space}
\date{}

\catcode`\@=11

\def\Let@{\relax\iffalse{\fi\let\\=\cr\iffalse}\fi}
\def\vspace@{\def\vspace##1{\crcr\noalign{\vskip##1\relax}}}
\def\multilimits@{\bgroup\vspace@\Let@
 \baselineskip\fontdimen10 \scriptfont\tw@
 \advance\baselineskip\fontdimen12 \scriptfont\tw@
 \lineskip\thr@@\fontdimen8 \scriptfont\thr@@
 \lineskiplimit\lineskip
 \vbox\bgroup\ialign\bgroup\hfil$\m@th\scriptstyle{##}$\hfil\crcr}
\def\Sb{_\multilimits@}
\def\endSb{\crcr\egroup\egroup\egroup}
\def\Sp{^\multilimits@}

\begin{document}
\maketitle

\begin{abstract}
The renormalized stress-energy tensor $\langle T_{\mu\nu}\rangle$ of the
quantized complex massless scalar field which obeys the automorphic
con\-di\-tion in Misner space is obtained. It is shown that there exists
the special value of the automorphic parameter for which $\langle
T_{\mu\nu}\rangle$ is regular on the chronology horizon and, so, can not
act as a protector of chronology through a back reaction on a
spacetime metric.  However, it is shown that, at the same time, the value
of field square $\langle\phi^2\rangle$, which characterizes the quantum
field fluctuati\-ons, is divergent on the chronology horizon.  The
assumption is sug\-ges\-ted that the infinitely growing quantum field
fluctuations, which appear if a (self)in\-ter\-action of the scalar field
is taken into account, would pre\-vent the chronology horizon formation.

\end{abstract}

\section{Introduction}

It has been known that classical general relativity permits spacetime to
develop closed timelike curves (CTCs) and, as a consequence, violate
causality (see \cite{Misner} and references therein). Believing in the
causality conservation, recently, Hawking has proposed a chronology
protection conjecture \cite{Hawking}, which states that the laws of
physics will always prevent the formation of closed timelike curves.
Hawking has argued that this conjecture might be fulfilled quite generally
due to vacuum polarization effects which will cause the expectation value
of the stress-energy tensor $\langle0|T_{\mu\nu}|0 \rangle$ to be
divergent at any chronology horizon\footnote{A chronology horizon ia a
surface in spacetime which separates a chronal region (i.e., a region that
contains no CTCs) from a non-chronal region (one containing CTCs through
every point). The chronology horizon is a special type of Cauchy horizon
(see for more details \cite{Misner}).} where CTCs are trying to form, and
will prevent the formation of CTCs through a back reaction on the
spacetime metric.

So far no proof of the chronology protection conjecture has been given.
So it seems important to study the vacuum polarization for different
physical fields on a chronology horizon in various spacetimes with CTCs.
In a number of previous papers [3-13] such investigation has been done.

In papers [3-11] it has been shown that $\langle0|T_{\mu\nu}|0 \rangle$
for a non-twisted massless scalar field diverges at the chronology horizon
in various spacetimes with CTCs. The back reaction of the metric to this
diverging stress-energy through the Einstein equations may be able to
prevent the formation of CTCs. However, in Boulware's work it has been
obtained that $\langle0|T_{\mu\nu}|0 \rangle$ for a massive scalar field
remains finite on the chronology horizon in Gott space \cite{Boulware}.
The similar result in Grant space was obtained by Tanaka and Hiscock. They
have found that $\langle0|T_{\mu\nu}|0 \rangle$ is finite on the
chronology horizon provided the mass of the scalar field is above a lower
limit which depends on the topological identification scale lengths of the
spacetime \cite{Tanaka2}. Thus, in these examples the metric backreaction
caused by a massive quantized field may not be large enough to
significantly change the space geometry and prevent the formation of CTCs.

One may see that the mass of a field could be a mechanism which provides
a regular behavior of $\langle0|T_{\mu\nu}|0 \rangle$ on the chronology
horizon. Another possibility has been pointed out by Frolov in
Ref.\cite{Frolov}. He has stressed that the sign  of the vacuum energy
density may depend on the spin of a field. In particular one may expect
that the contribution of fermions to the energy density has an opposite
sign than the contribution of bosons. And, in principle, the situation
could be when there is an exact cancellation of the leading contributions
of all fields (as it happens for the vacuum energy density in flat
spacetime in a supersymmetric theory). In the later case
$\langle0|T_{\mu\nu}|0 \rangle$ could be finite on the chronology
horizon.

In the previous paper \cite{SushkovTMF} I have considered
two-dimensional model of ''time machine'' with an automorphic complex
scalar field. The automorphic fields $\phi(X)$ ($X$ is a spacetime point)
are those ones which obey the generalized periodic (or automorphic)
condition $\phi(\gamma X) = a(\gamma)\phi(X)$, where $a^2(\gamma)=1$, and
$\gamma$ are elements of the discrete group of isometry $\Gamma$ on a
spacetime; operators $\gamma$ act in the following way:  $\gamma
X=\widetilde X$, where points $X$ and $\widetilde X$ are identified. Note
that the Lagrangian ${\cal L}[\phi(X)]$ of free fields is quadratic in
$\phi $ so it is invariant under the symmetry transformations $\gamma$ of
spacetime. In the case of the complex scalar field the automorphic
condition takes the form
\begin{equation} \label{auto}
\phi (\gamma X)=e^{2\pi i\alpha}\phi(X)\,,\quad
0\leq\alpha\leq\frac{1}{2}\,,
\end{equation}
where $\alpha$ is an automorphic parameter. There are two particular cases
$\alpha=0$ and $\alpha=\frac{1}{2}$ for which the condition (\ref{auto})
reads $\phi (\gamma X)=\phi (X)$ (a non-twisted field) and \mbox{$\phi
(\gamma X)=-\phi (X)$} (a twisted field), respectively. (For more
details about automorphic fields, see Refs. 15 and 16.)

In the paper \cite{SushkovTMF} it has been shown that
$\langle0|T_{\mu\nu}|0 \rangle$ for the complex scalar field remains
finite on the chronology horizon if the automorphic parameter has some
specific values. This result is not a surprise. Moreover, one may expect
to obtain the similar results for automorphic fields in any spacetimes
with CTCs. Really, it is known that the sign of the vacuum energy density
is different for the non-twisted ($\alpha=0$) field and the twisted
($\alpha=\frac{1}{2}$) one. So one may suppose that near the chronology
horizon a diverging part of $\langle0|T_{\mu\nu}|0 \rangle$ will have
different signs for cases of a non-twisted field and twisted one. And one
may also expect that in the case of some intermediate value of the
automorphic parameter $\alpha$ the diverging part of
$\langle0|T_{\mu\nu}|0 \rangle$ will vanish, so that the vacuum
expectation values of $\langle0|T_{\mu\nu}|0 \rangle$ will be finite on
the chronology horizon.

In this paper we continue an investigation of a quantized complex scalar
field in a spacetime with CTCs in order to understand better a role of
automorphic fields in a mechanism of the chronology protection. In
previous work \cite{SushkovTMF} the two-dimensional particular model of a
time machine has been considered. Here we shall consider the Misner space
which is convenient for next reasons: (i) Misner space is a flat spacetime
with non-trivial topology so it does not need matter fields with the
negative energy density as a wormhole spacetime does. (ii) Misner space
has a simple mathematical structure so it becomes possible to carry out
the exact calculation of $\langle0|T_{\mu\nu}|0 \rangle$. This calculation
is interesting itself from the viewpoint of quantum field theory on a flat
space with a non-trivial topology.

In section 2 we briefly review the basic structure of the Misner space.
The vacuum stress-energy tensor for a complex scalar field is computed in
section 3, and its behavior near the chronology horizon is analysed in
section 4. The units $c=\hbar=G=1$ are used through the paper.

\section{Misner space}

In details the properties of Misner space have been analysed in Refs.[2,
10, 13]. Here we give a brief review of this properties.

Misner space was originally developed to illustrate topological
patho\-logies associated with Taub-NUT (Newman-Unti-Tamburino) type
spa\-ce\-times [17,18]. It is simply the flat Kasner universe with
$S^1\times R^3$ topology. In the Misner coordinates $(t,x^1,x^2,x^3)$ the
metric is given by
\beq\label{Misnermetric}
ds^2=-dt^2+t^2(dx^1)^2+(dx^2)^2+(dx^3)^2\,.
\eeq
By transforming to a new set of coordinates $\{ y^{\alpha}\}$ defined by
\beq\label{coordrelation}
y^0=t\cosh(x^1)\,,\quad y^1=t\sinh(x^1)\,,\quad y^2=x^2\,,\quad y^3=x^3\,,
\eeq
the Misner space metric becomes identical with the Minkowski space metric,
\beq
ds^2=-(dy^0)^2+(dy^1)^2+(dy^2)^2+(dy^3)^2\,.
\eeq
The unique properties of Misner space originate in its topology. In the
Misner coordinates, the spacetime is taken to be periodic in the $x^1$
direction with period $a$. Thus, the following points are identified
with one another:
\beq
(t,x^1,x^2,x^3)\leftrightarrow (t,x^1+na,x^2,x^3)\,.
\eeq
In the Minkowski coordinates, the above topological identification takes
the form
%\beq\label{ident}
\begin{eqnarray}\label{ident}
\lefteqn{(y^0,y^1,y^2,y^3)\leftrightarrow} \nonumber\\
& & (y^0\cosh(na)+y^1\sinh(na),\,y^0\sinh(na)+y^1\cosh(na),\,y^2,y^3)\,.
\end{eqnarray}
%\eeq
 From (\ref{ident}) one may see that the Misner space identifications are
equivalent to a discret Lorentz boost by a velocity equal to $\tanh(na)$
in the $y^1$ direction in the Minkowski space.

Consider the $t$-$x^1$ plane of the  Misner space and the light cone in
the two-dimensional Minkowski space. The lower half-plane ($t<0$) of the
Misner space is mapped onto the lower quadrant (the past light cone) of
the Minkowski space. The Misner metric (\ref{Misnermetric}) has an
apparent singularity at $t=0$. However, one can extend it by introducing
new coordinates
\begin{displaymath}
\tau=t^2\,,v=\ln t+x^1\,.
\end{displaymath}
The metric then takes the form
\begin{displaymath}
ds^2=-dvd\tau+\tau dv^2\,.
\end{displaymath}
This can then be extended through $\tau=0$. This corresponds to extending
from the lower quadrant into the left-hand quadrant. However, at $\tau=0$,
the light cone tip over and a closed null geodesic appears. For negative
$\tau$, closed timelike curves appear. The same result is obtained for the
upper half-plane ($t>0$) of the Misner space and the future light cone in
the Minkowski space. Thus, the $t=0$ surface in the extended Misner space
separates the region with CTCs from the region without them. This
surface is a chronology horizon.

\section{$\langle T_{\mu\nu}\rangle^{ren}$ and
$\langle\phi^2\rangle^{ren}$ for complex scalar field in Misner
space}
Consider a complex massless scalar field $\phi$ with
the stress-energy tensor
\begin{eqnarray}
     \lefteqn{T_{\mu\nu}=(1-2\xi)
     \nabla_{(\mu}\phi\nabla_{\nu)}\bar\phi+(4\xi-1)g_{\mu\nu}
     \nabla_{\alpha}\phi\nabla^{\alpha}\bar\phi-}\nonumber\\
& &  2\xi(\phi\nabla_{\mu}\nabla_{\nu}\bar\phi+
     \bar\phi\nabla_{\mu}\nabla_{\nu}\phi)
     +\frac{1}{2}\xi g_{\mu\nu}(\phi \Box\bar\phi+\bar\phi\Box\phi)\,.
\end{eqnarray}
The scalar field $\phi$ satisfies the field equation
\beq\Box\phi=0\,.\eeq
(The complex conjugate field obeys the same equation $\Box\bar\phi=0$.)
In the Minkowski coordinates $\{y^{\alpha}\}$, a general
positive-frequency solution of this equation can easily be written in the
form
\begin{eqnarray}\label{gensol}
     \lefteqn{\phi(y^0,y^1,y^2,y^3)=}\nonumber\\
& &  \int\!\!\!\int\!\!\!\int
     \frac{dk_1dk_2dk_3A(k_0,k_1,k_2,k_3)}
     {4\pi^{3/2}k_0}e^{i(-k_0y^0+k_1y^1+k_2y^2+k_3y^3)}\,,
\end{eqnarray}
where $k_0=\sqrt{k_1^2+k_2^2+k_3^2}$, and $A(k_0,k_1,k_2,k_3)$ is an
arbitrary ''spec\-tral'' function. Now let us demand that this solutions
obey the automor\-phic condition (\ref{auto}), which in Misner space takes
the form
\begin{displaymath}
\phi(y^0\cosh a+y^1\sinh a,y^0\sinh a+y^1\cosh a,y^2,y^3)=
\end{displaymath}
\beq
e^{2\pi i\alpha}\phi(y^0,y^1,y^2,y^3)\,.
\eeq
Positive-frequency solutions (\ref{gensol}) will be automorphic provided
\begin{displaymath}
A(k_0\cosh a-k_1\sinh a,k_1\cosh a-k_0\sinh a,k_2,k_3)=
\end{displaymath}
\beq
e^{-2\pi i\alpha}A(k_0,k_1,k_2,k_3)\,.
\eeq
The general solution of this functional equation is
\beq
A(k_0,k_1,k_2,k_3)=\sum_{n}C_{n}(k_2,k_3)(k_0-k_1)^{i\nu}\,,
\eeq
where $\nu=-2\pi a^{-1}(n+\alpha)$, $C_{n}(k_2,k_3)$ are arbitrary
functions of $k_2$ and $k_3$, and the summation is taken over all integer
numbers $n$. Substituting the expression for $A(k_0,k_1,k_2,k_3)$ in
(\ref{gensol}) and carrying out the integration over $k_1$ one can obtain the
following representation for a general positive-frequency automorphic
solution:
\beq\label{sol}
\phi(y^0,y^1,y^2,y^3)=\sum_{n}\int\!\!\!\int dk_2dk_3
\widetilde{C}_{n}(k_2,k_3)\phi_J(y^0,y^1,y^2,y^3)\,,
\eeq
where $J$ is a multi-index $\{ n,k_2,k_3\}$, and modes
$\phi_J(y^0,y^1,y^2,y^3)$ is given by
%\beq\label{autosol}
\begin{displaymath}
     \phi_J(y^0,y^1,y^2,y^3)=D_n(y^0+y^1)^{i\nu/2}(y^0-y^1)^{-i\nu/2}
     \times
\end{displaymath}
\beq\label{autosol}
     H_{i\nu}^{(2)}\left(\kappa\sqrt{(y^0)^2-(y^1)^2}\right)
     e^{i(k_2y^2+k_3y^3)}\,
\eeq
$\kappa=\sqrt{(k_2)^2+(k_3)^2}$, and $D_n$ are normalizing coefficients.
Below it will be convenient to use the Misner coordinate set
$\{X^{\alpha}\}=(t,x^1,x^2,x^3)$. Using the relation (\ref{coordrelation})
between Minkowski and Misner coordinates one can rewrite the solutions
(\ref{autosol}) as follows:
\beq\label{solutions}
\phi_J(t,x^1,x^2,x^3)=D_nH_{i\nu}^{(2)}(\kappa t)
e^{i(\nu x^1+k_2x^2+k_3x^3)}\,.
\eeq
Let us introduce the following scalar product
$\langle\phi_1,\phi_2\rangle$ in the space of automorphic solutions
(\ref{sol})
\beq
\langle\phi_1,\phi_2\rangle=i\int_{\Sigma}
[\phi_1(X)\partial_{\mu}{\bar\phi}_2(X)-
{\bar\phi}_2(X)\partial_{\mu}\phi_1(X)]d\sigma^{\mu}\,,
\eeq
where $d\sigma^{\mu}$ is a surface element of a Cauchy surface $\Sigma$ in
Misner space. As the value of this scalar product does not depend on the
particular choice of $\Sigma$, one may choose the surface $\Sigma$ defined
by the equation $t=$constant. Taking into account the periodicity of
Misner space in $x^1$ direction, one may represent the scalar product as
follows:
\beq\label{scproduct}
\langle\phi_1,\phi_2\rangle=-it\int_{0}^{a}\!\!\!
\int\!\!\!\int_{t=const}
[\phi_1(X)\partial_{t}{\bar\phi}_2(X)-
{\bar\phi}_2(X)\partial_{t}\phi_1(X)]dx^1dx^2dx^3\,.
\eeq
The set of positive-frequency automorphic solutions (\ref{sol}) with the
scalar product (\ref{scproduct}) forms a Hilbert space $H$. The solutions
(\ref{solutions}) form an orthonormal basis in $H$:
$\langle\phi_n,\phi_m\rangle=\delta_{nm}$ provided
\beq
D_n=\frac{e^{\pi\nu/2}}{4\sqrt{\pi a}}\,.
\eeq
The Hadamard function $G^{(1)}(X,\widetilde{X})$ is defined by
\beq\label{defhadamard}
G^{(1)}(X,\widetilde{X})=\sum_J[\phi_J(X)\bar\phi_J(\widetilde{X})+
\phi_J(\widetilde{X})\bar\phi_J(X)]\,,
\eeq
where $\sum_J=\sum_n\int\!\!\!\int dk_2dk_3$. Substituting the solutions
(\ref{solutions}) into the last expression, one can obtain after
calculations the following expression for the Hadamard function $G^{(1)}$:
\begin{eqnarray}\label{lastHad}
    \lefteqn{G^{(1)}(X,\widetilde{X})=\frac{1}{4\pi^2\sigma}+}\nonumber\\
& & \frac{1}{\pi^2t\tilde t\sqrt{1-\chi^2}}\sum_{n=1}^{\infty}
    \cosh[n(x^1-\tilde x^1)]\sin(n\arccos\chi)\Psi_n(\alpha,a)\,,
\end{eqnarray}
where
\beq
\sigma=\frac{1}{2}\left[-(t-\tilde t)^2+2t\tilde t(\cosh(x^1-\tilde x^1)-1)
+(x^2-\tilde x^2)^2+(x^3-\tilde x^3)^2\right]\,,
\eeq
and
\beq
\Psi_n(\alpha,a)=\frac{e^{na}\cos2\pi\alpha-1}
{e^{2na}-2e^{na}\cos2\pi\alpha+1}
\eeq
To obtain the renormalized Hadamard function for the Misner space we must
subtract from the expression (\ref{lastHad}) the divergent Minkowski
vacuum state term
\beq
G^{(1)}_0(X,\widetilde X)=\frac{1}{4\pi^2\sigma_0}\,,
\eeq
where $\sigma_0=\frac{1}{2}g_{\alpha\beta}(X^{\alpha}-\widetilde
X^{\alpha}) (X^{\beta}-\widetilde X^{\beta})$ is a geodetic interval,
which in the Misner coordinates is given by
\beq
\sigma_0=\frac{1}{2}[-(t-\tilde t)^2+t^2(x^1-\tilde x^1)^2
+(x^2-\tilde x^2)^2+(x^3-\tilde x^3)^2]\,.
\eeq
So, the renormalized Hadamard function $G^{(1)}_{ren}(X,\widetilde X)$ is
\begin{eqnarray}\label{renHad}
    \lefteqn{G^{(1)}_{ren}(X,\widetilde{X})=\frac{1}{4\pi^2}
    \left( {1\over\sigma}-{1\over\sigma_0}\right)+}\nonumber\\
& & \frac{1}{\pi^2t\tilde t\sqrt{1-\chi^2}}\sum_{n=1}^{\infty}
    \cosh[n(x^1-\tilde x^1)]\sin(n\arccos\chi)\Psi_n(\alpha,a)\,.
\end{eqnarray}
Now we may find the renormalized vacuum expectation values of the
stress-energy tensor $\langle0|T_{\mu\nu}|0 \rangle$ and the field square
$\langle0| \phi^2|0 \rangle$ (hereafter $\langle T_{\mu\nu}\rangle$
and $\langle \phi^2\rangle$).

The value of $\langle \phi^2\rangle$, which characterize the vacuum
fluctuations, is defined as follows:
\beq
\langle\phi^2\rangle=\lim_{\widetilde X\to X}G^{(1)}_{ren}
(X,\widetilde X)\,,
\eeq
Hence, using the expression (\ref{renHad}) for $G^{(1)}_{ren}$,
we can obtain
\beq\label{phi}
     \langle\phi^2\rangle=\frac{K}{t^2}\,,\quad
     {\textstyle \rm where}\quad K\equiv K(\alpha,a)=
     \frac{1}{\pi^2}\sum_{n=1}^{\infty} n\Psi_n(\alpha,a)\,.
\eeq

The vacuum expectation value of the stress-energy tensor for a complex
massless scalar field can be written as
\begin{displaymath}
\langle T_{\mu\nu}\rangle=\lim_{\widetilde X\to X}\left\{(1-\xi)
\nabla_{\mu}{\widetilde\nabla}_{\nu}+(2\xi-{\textstyle
{1\over2}})g_{\mu\nu} \nabla_{\alpha}{\widetilde\nabla}^{\alpha} \right.
\end{displaymath}
\beq
\left. -2\xi\nabla_{\mu}\nabla_{\nu}+{\textstyle {1\over2}}\xi
g_{\mu\nu}\Box\right\}G^{(1)}_{ren}(X,\widetilde X)\,.
\eeq
Using the expression (\ref{renHad}) for $G^{(1)}_{ren}$, computing the
derivatives and taking the limit as the separated points are brought
together, one can obtain the following expressions for the components of
$\langle T_{\mu\nu}\rangle$ with arbitrary curvature coupling:
\beq\label{SETxi}
\langle T_{\mu\nu}\rangle=\frac{1}{t^4}{\textstyle \rm
diag}(L,3L,M,M)\,,
\eeq
where
\beq
L\equiv L(\xi,\alpha,a)=\frac{1}{\pi^2}\sum_{n=1}^{\infty}
\left(\frac{n-n^3}{3}-\frac{3\xi n}{2}\right) \Psi_n(\alpha,a)\,,
\eeq
\beq
M\equiv M(\xi,\alpha,a)=\frac{1}{\pi^2}\sum_{n=1}^{\infty}
\left(\frac{2n+n^3}{3}-\frac{9\xi n}{2}\right) \Psi_n(\alpha,a)\,.
\eeq
For conformal coupling $(\xi=\frac{1}{6})$ the stress-energy tensor
$\langle T_{\mu\nu}\rangle$ has the form
\beq\label{SET}
        \langle T_{\mu\nu}\rangle=\frac{1}{t^4}{\textstyle \rm
        diag}(N,3N,-N,-N)\,,
\eeq
where
\begin{displaymath}
     N\equiv N(\alpha,a)=L({\textstyle\frac{1}{6}},\alpha,a)
     =-M({\textstyle\frac{1}{6}},\alpha,a)=
\end{displaymath}
\beq
     \frac{1}{\pi^2}\sum_{n=1}^{\infty}
     \left(\frac{n}{12}-\frac{n^3}{3}\right) \Psi_n(\alpha,a)\,.
\eeq

\section{Behavior of $\langle T_{\mu\nu}\rangle$ and
$\langle\phi^2\rangle$ near the chro\-no\-logy horizon}

As has been shown in section 2 the chronology horizon in Misner space
is the $t=0$ surface. One may see that the expressions (\ref{phi}),
(\ref{SETxi}) and (\ref{SET}) for $\langle\phi^2\rangle$ and
$\langle T_{\mu\nu}\rangle$ are generally diverged on the chronology
horizon, when $t$ goes to zero. However, it may be possible that the
factors $K$, $L$, $M$, or $N$ become equal to zero. In this case
the corresponding components of $\langle T_{\mu\nu}\rangle$ and/or
$\langle\phi^2\rangle$ are equal to zero too and they remain regular on
the chronology horizon.

Consider the case of non-conformal coupling $(\xi\not=\frac{1}{6})$. In
this case $\langle T_{\mu\nu}\rangle$ is given by the expression
(\ref{SETxi}). Coefficients $L$ and $M$ are the functions
$L(\xi,\alpha,a)$ and $M(\xi,\alpha,a)$ of parameters $\xi$, $\alpha$
(automorphic parameter) and $a$ (periodicity parameter). Let us fix two
parameters $\xi$ and $a$.  The graphs of $L$ and $M$ as functions of
$\alpha$ is shown in the figure 1. One may see that the functions $L$ and
$M$ have alternating signs. They become equal to zero at some values
of $\alpha$, which are different for $L$ and $M$, i.e. the coefficients
$L$ and $M$ are not equal to zero simultaneously. Note that this behavior
of the functions $L$ and $M$ remains qualitatively the same for all values
of $\xi$ and $a$. Thus, we can do the conclusion that the components of
$\langle T_{\mu\nu}\rangle$ (at least, the some of them) diverge at
the chronology horizon for all values $\xi\not=\frac{1}{6}$, $\alpha$ and
$a$.
\begin{figure}[hl]
\vspace{4cm}
\caption{Graphs of $L(\xi,\alpha,a)$ (solid line) and $M(\xi,\alpha,a)$
(dashed line) as functions of $\alpha$.}
\end{figure}

Now consider the case of conformal coupling $\xi=\frac{1}{6}$. In
this case $\langle T_{\mu\nu}\rangle$ is given by the expression
(\ref{SET}). We may see that the behavior of $\langle T_{\mu\nu}\rangle$
is only determined by the coefficient $N$. The family of the graphs of
$N(\alpha,a)$ as a function of $\alpha$, corresponding to the various
values of $a$, is shown in the figure 2 (the solid lines). From figure 2
one can see the next characteristic features of the behavior of $N$. The
function $N(\alpha,a)$ has alternating sings, and, for any value of $a$,
it becomes zero at the point $$\alpha=\alpha_{*}\approx 0.24\,.$$

We obtain that the coefficient $N$ is equal to zero if $\alpha=\alpha_{*}$.
But if $N=0$, then all components of $\langle T_{\mu\nu}\rangle$ are
identically equal to zero too. Thus, we can conclude that there exists
the special case with the automorphic parameter $\alpha=\alpha_{*}$, when
the stress-energy tensor $\langle T_{\mu\nu}\rangle$ of the complex
massless scalar field vanishes in the whole Misner space.
Of course, this ''everywhere null'' stress-energy tensor is
also equal to zero on the chronology horizon, and so it can not prevent
closed timelike curves from appearing through the back reaction on the
spacetime metric.

Does it means that this field configuration can not act as a protector of
chronology? To answer, let us consider the behavior of $\langle\phi^2
\rangle$. $\langle\phi^2\rangle$ is given by the expression
(\ref{phi}) and fully determined by the coefficient $K$. The family of
the graphs of $K(\alpha,a)$ as a function of $\alpha$, corresponding to
the various values of $a$, is shown in the figure 2 (the dashed lines).
One may see that $K$ is a function with alternating signs, which becomes
zero at some point $\alpha=\alpha_{\star}$. It is important that
$\alpha_{\star}\not=\alpha_{*}$. Hence, when the components of
$\langle T_{\mu\nu}\rangle$ are equal to zero, at the same time
$\langle\phi^2\rangle$ is distinct from zero. Thus, we may conclude that
$\langle\phi^2\rangle$ will be divergent on the chronology horizon, and, as
a consequence, the vacuum field fluctuations will be infinitely growing
as the chronology horizon try to form.
\begin{figure}[hl]
\vspace{4cm}
%\special{fried1.}
\caption{Family of graphs of $K(\alpha,a)$ (dashed lines) and $N(\alpha,a)$
(solid lines) as functions of $\alpha$.}
\end{figure}
This infinitely growing vacuum fluctuations may
lead to an appear\-ance of the infinite vacuum energy density if a
(self)interaction of the scalar field is taken into account. As real
physical fields are non-free, one may suppose that the vacuum fluctuations
could be a mechanism that prevents the chronology horizon from forming.

\end{document}